\documentclass[fleqn,12pt,twoside]{article}
\usepackage{espcrc1}
 
\usepackage{graphicx}
\usepackage[figuresright]{rotating}
 
        \newcommand{\be}{\begin{equation}}
        \newcommand{\ee}{\end{equation}}
        \newcommand{\bea}{\begin{eqnarray}}
        \newcommand{\eea}{\end{eqnarray}}

\title{The $\theta$-Vacua and the Leutwyler--Smilga Scaling Regime}                  

\author{J.T.\ Lenaghan\address{The Niels Bohr Institute, Blegdamsvej 17, DK-2100 \\
        Copenhagen \O, Denmark \\
        email: lenaghan@alf.nbi.dk}%
        \thanks{Work done in collaboration with G.\ Akemann, K.\ Splittorff and T.\ Wilke.}       }                 
\begin{document}
 
\maketitle
 
\begin{abstract}
The partition function of QCD is studied in the Leutwyler--Smilga
scaling regime for an arbitrary number of quark flavors and masses
including the contributions from all winding numbers.  For $N_f=2$ and
degenerate quark masses, the partition function becomes independent of
the quark masses at $\theta=\pi$ and subsequently the scalar chiral
condensate vanishes.  There is a discontinuity at $\theta=\pi$ in the
first derivative of the energy density with respect to $\theta$
corresponding to the first--order phase transition in which CP is
spontaneously broken, known as Dashen's phenomena.  These
properties are found to be insensitive to both the pattern of chiral
symmetry breaking and the representation of the quark fields.
\end{abstract}            

\section{Introduction}

The vacuum angle, $\theta$, a fundamental parameter in QCD, is
constrained to be zero to within one part in a trillion.  
Understanding why this parameter is so fine--tuned is a fascinating
theoretical problem.  Dashen's phenomenon \cite{Dashen}, the
first--order phase transition at $\theta=\pi$ in which the discrete CP
symmetry is spontaneously broken, is an example of novel physics at
nonzero values of $\theta$.  Since the physics at nonzero $\theta$ is
inherently nonperturbative, the only two theoretical avenues to
explore this region of the QCD parameter space are lattice QCD and
effective theories.  The former approach suffers from a sign--problem
which is similar to that which plagues attempts to study QCD at 
nonzero baryonic chemical potential.  

The approach employed here is to study the physics of nonzero $\theta$
using chiral perturbation theory in finite volume \cite{Gasser}.
Restricting the Euclidean four-volume, $V=L^4$, to the range
$\frac{1}{\Lambda} \ll L \ll \frac{1}{m_\pi}$ where $\Lambda$ is the
chiral symmetry breaking scale and $m_\pi$ is the mass of the Goldstone
excitations, allows for an {\em exact} analytical treatment
\cite{Leutwyler:1992yt}.  This is possible since the lower limit
implies that the partition function is dominated by Goldstone modes
and the upper limit requires that these modes are constant.
As a result, the partition function reduces to a finite--dimensional 
group integration.

\section{Results}
The full partition function for $N_f=2$ and $N_c\ge3$ with quarks in
the fundamental representation was calculated in Ref.\
\cite{Leutwyler:1992yt} and the necessity of including the
contributions from all topological sectors in the partition function
was demonstrated in Ref.\ \cite{Damgaard1}.  
For $N_f \ge 3$, the
problem is technically more difficult.  It was shown in Ref.\
\cite{lw} that the full partition function for arbitrary $N_f$,
$\theta$ and quark masses, $m_i$, can be expressed as a $N_f-2$ dimensional
integral over single Bessel functions.

For simplicity, we focus on $N_f=2$ and $N_c\ge3$ with quarks
in the fundamental representation but the following results 
are generalizable \cite{lw,ALS}.
The partition function is ${\cal Z}(\theta,\mu_1,\mu_2)=
\frac{I_1\left(\mu_{12}\right)}{\mu_{12}}$ where $\mu_{12} =
\sqrt{\mu_1^2+\mu_2^2+2\mu_1 \mu_2 \cos(\theta)}$,
$\mu_i=\Sigma\, V\, m_i$ and $\Sigma$ is 
the infinite-volume chiral condensate at $\theta=0$ \cite{Leutwyler:1992yt}.  
Taking the two quark masses to be equal and $\theta$ to be an integer
multiple of $\pi$
leads to the remarkable property that the partition function is {\em
independent} of the scaling variable with ${\cal
Z}^{(N_f=2)}(\theta=n\pi,\mu)=\frac{1}{2}$.  
For arbitrary $\theta$ and degenerate quark masses, 
the dependence on the volume, macroscopic chiral condensate,
quark masses and $\theta$ is only through the dimensionless variable
$\mu |\cos(\theta/2)|$, which as discussed in Ref.\ \cite{ALS}, has
profound implications.  On account of this very constrained 
dependence of the partition function on the parameters of the theory, 
the chiral and the topological properties are intimately intertwined. 
In particular, the chiral condensate and the topological density are 
proportional up to a simple $\theta$-dependent function \cite{ALS}.

Since the analytic form of the partition function 
is known, calculating the vacuum properties is straightforward. 
Taking the limit of degenerate quark masses, the chiral condensate is 
$\Sigma(\theta,\mu)= \Sigma\,|\cos(\theta/2)|  \, 
I_2\left(2 \mu |\cos(\theta/2)|\right)/\left[2 I_1\left(2 \mu |\cos(\theta/2)|\right)\right]$.
For fixed $\mu$, the chiral condensate decreases monotonically in the
interval $\theta\in [0,\pi)$ and increases monotonically for $\theta
\in (\pi,2\pi)$.  As $\mu=m \, \Sigma \, V \rightarrow \infty$, the
chiral condensate develops a cusp at $\theta=\pi$.  For any value of
$\mu$, the chiral condensate vanishes identically at $\theta=\pi$.  In
the limit of degenerate quark masses, the topological density, defined
as the first derivative of the logarithm of the partition function
with respect to $\theta$, is 
$\sigma(\theta,\mu) = m \tan(\theta/2) \Sigma(\theta,\mu)$.  
The most interesting property of this relation is
that in the limit of very large scaling variable $\sigma(\theta,\mu)$
develops a discontinuity at $\theta=\pi$.  This is the
first--order phase transition proposed by Dashen \cite{Dashen}.  The
topological susceptibility can also be calculated with the infinite
volume result $\chi(\theta,m,V)=\frac{\Sigma
m}{2}|\cos(\theta/2)|$.  This is nothing but the flavor singlet
Ward--Takahashi identity generalized to nonzero values of $\theta$
which predicts a linear rise in the topological susceptibility with
the quark mass.
All of the above results can
be shown to be independent of both the pattern of chiral
symmetry breaking and the representation of the matter fields
\cite{ALS}.

\end{document}